\documentclass[twocolumn,showkeys,aps,prb,showpacs]{revtex4-1}
\UseRawInputEncoding
\usepackage{graphicx}
\usepackage[CJKbookmarks,dvipdfm,colorlinks,linkcolor=red,citecolor=blue]{hyperref}

\begin{document}

\title{ Valley polarization transition driven  by biaxial strain in Janus  $\mathrm{GdClF}$ monolayer}

\author{San-Dong Guo$^{1}$, Xiao-Shu Guo$^{1}$,  Xiu-Xia Cai$^{1}$  and  Bang-Gui Liu$^{2,3}$}
\affiliation{$^1$School of Electronic Engineering, Xi'an University of Posts and Telecommunications, Xi'an 710121, China}
\affiliation{$^2$ Beijing National Laboratory for Condensed Matter Physics, Institute of Physics, Chinese Academy of Sciences, Beijing 100190, People's Republic of China}
\affiliation{$^3$School of Physical Sciences, University of Chinese Academy of Sciences, Beijing 100190, People's Republic of China}
\begin{abstract}
The valley degrees of freedom of carriers in crystals is useful to process information and perform logic operations, and it is a key factor for valley application  to realize the valley polarization. Here, we propose a model that  the valley polarization transition at different valley points (-K and K points) is produced by biaxial strain. By the first-principle calculations, we illustrate our idea with a concrete example of Janus  $\mathrm{GdClF}$ monolayer.
The predicted  $\mathrm{GdClF}$ monolayer is dynamically, mechanically  and thermally  stable,  and is a  ferromagnetic (FM) semiconductor with  perpendicular magnetic anisotropy (PMA), valence band maximum (VBM) at valley points and   high Curie temperature ($T_C$).
Due to its intrinsic ferromagnetism and  spin orbital
coupling (SOC),  a  spontaneous valley polarization will be induced, but the valley splitting is only -3.1 meV, which provides an opportunity to achieve valley polarization transition at different valley points by strain. In considered strain range ($a/a_0$: 0.94$\sim$1.06), the strained GdClF monolayer has always energy bandgap,  strong FM coupling  and PMA.  The compressive strain is in favour of -K valley polarization, while the tensile strain makes for K valley polarization. The corresponding valley splitting  at 0.96 and 1.04 strain are -44.5 meV and 29.4 meV, which are higher than
the thermal energy of  room temperature (25 meV). Due to special Janus structure, both  in-plane and  out-of-plane piezoelectric polarizations can be observed.  It is found that the direction of in-plane piezoelectric polarizations can be overturned by strain, and the $d_{11}$  at 0.96 and 1.04 strain
are -1.37 pm/V and 2.05 pm/V.
 Our works  pave the way to design the ferrovalley material as multifunctional
valleytronics and piezoelectric  devices by strain.

\end{abstract}
\keywords{Valleytronics, Ferromagnetism, Piezoelectronics, 2D materials}

\pacs{71.20.-b, 77.65.-j, 72.15.Jf, 78.67.-n ~~~~~~~~~~~~~~~~~~~~~~~~~~~~~~~~~~~Email:sandongyuwang@163.com}

\maketitle

\section{Introduction}
In analogy to charge and spin
of electrons,  valley degree of freedom  constitutes the binary
states in crystals, which  lays the foundation for
 processing information and performing logic operations\cite{q1,q2,q3,q4,q5,q6,q9}. The valley-related filed is truly flourishing with the advent of two-dimensional (2D) materials.
 Valley is referred to two or more local energy extremes in
the conduction band or valence band. These valleys are degenerate but inequivalent, which  are
well separated in the 2D momentum space with
suppressed intervalley interactions\cite{q1}. Nevertheless, to achieve valley application, it  is very  necessary to realize the valley polarization.
The optical pumping, magnetic field, magnetic
substrates and  magnetic doping have been proposed to induce polarization\cite{q9-1,q9-2,q9-3,q9-4}. However,  these methods may destroy
the  intrinsic  energy band structures and crystal structures of these valley materials. Then, the  2D magnetic materials provide an opportunity to
realize spontaneous valley polarization (namely ferrovalley  materials)\cite{q10}, and many 2D materials have been predicted with  spontaneous valley polarization, such as  2H-$\mathrm{VSe_2}$\cite{q10}, $\mathrm{CrSi_2X_4}$ (X=N and P)\cite{q11}, $\mathrm{VAgP_2Se_6}$\cite{q12}, $\mathrm{LaBr_2}$\cite{q13,q13-1}, $\mathrm{VSi_2P_4}$\cite{q14},  $\mathrm{NbX_2}$ (X =S and Se)\cite{q15},
$\mathrm{Nb_3I_8}$\cite{q16}, $\mathrm{TiVI_6}$\cite{q17},  FeClBr\cite{q18}.

\begin{figure}
  \includegraphics[width=8.0cm]{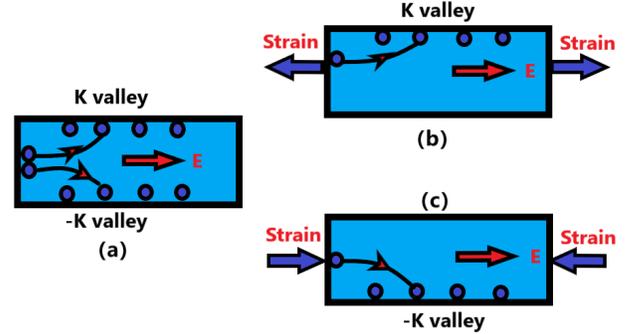}
  \caption{(Color online)Sketch of strain-induced  anomalous valley Hall effect under an in-plane longitudinal electric field $E$. (a)Without strain, no valley polarization is produced. (b)With applied tensile strain, the K valley is  polarized. (c)
  With applied compressive strain, the -K valley is polarized. }\label{v-0}
\end{figure}
\begin{figure*}
  \includegraphics[width=13.6cm]{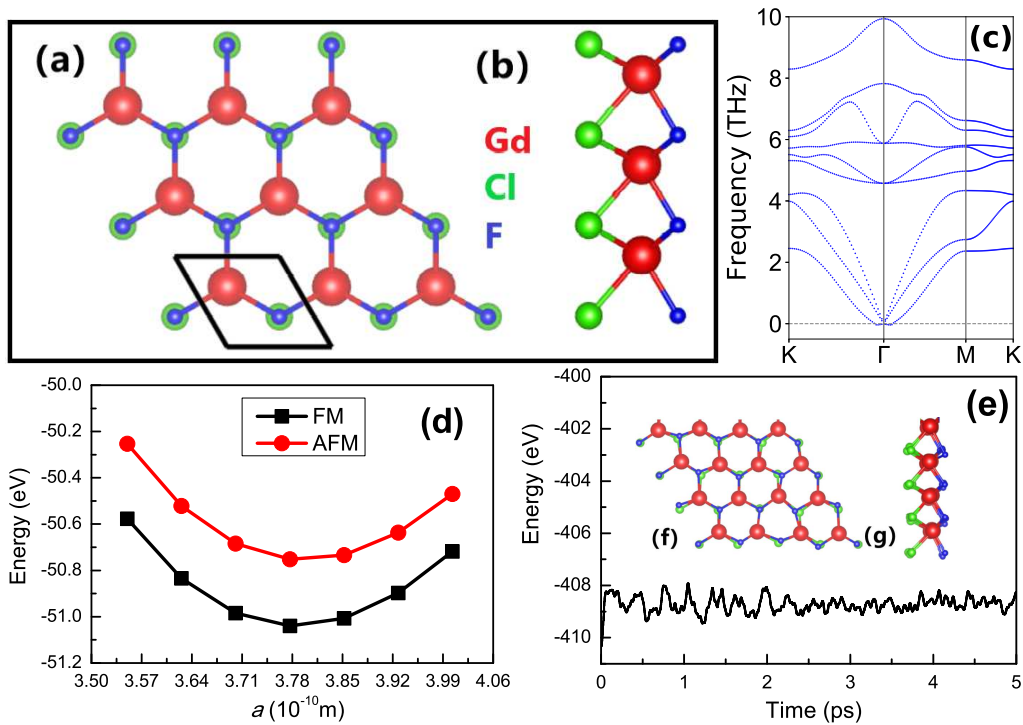}
  \caption{(Color online)The (a) top view and (b) side view of  crystal structure of Janus  monolayer  $\mathrm{GdClF}$, and the
   black frame represents the rhombus primitive cell. (c) The  phonon dispersions   of  Janus monolayer  $\mathrm{GdClF}$ using GGA+U. (d) The FM and AFM  energies  of Janus  monolayer  $\mathrm{GdClF}$ as a function of  lattice constants $a$  with rectangle supercell. (e) The variation of free energy  of Janus  monolayer  $\mathrm{GdClF}$ as a function of  simulation time. Insets show the
 final structures  of Janus  monolayer  $\mathrm{GdClF}$ after 5 ps at 300 K (top view (f) and side view (g)).}\label{st}
\end{figure*}
Although a 2D material is ferrovalley  material, its valley splitting may be  zero or very little due to accidental factor. Taking 2D hexagonal lattice for example, when compressive/tensile strain is applied, -K/K valley will be polarized.  When an in-plane longitudinal electric field $E$ is imposed,
the Bloch electrons  will  acquire an anomalous Hall velocity
$\upsilon$  due to $\upsilon\sim E\times\Omega(k)$\cite{q9},  where the $\Omega(k)$ is Berry curvature.   Then, the anomalous valley Hall effect (AVHE) will be produced by the carriers of  -K/K valley with compressive/tensile strain. The proposed model is
illustrated in \autoref{v-0}.
Recently, a kind of f-electrons 2D FM semiconductors  $\mathrm{GdX_2}$ (X=F, Cl, Br and I) are reported, which possess  high  Curie temperature\cite{g0,g1,g2}.  The monolayer $\mathrm{GdI_2}$ of them is predicted as a ferrovalley  material with a giant splitting of 149 meV\cite{g3}.  However, $\mathrm{GdI_2}$  possesses in-plane magnetic anisotropy, not PMA\cite{g1,g2}.
Because the pace group of $\mathrm{GdX_2}$ (X=F, Cl, Br and I) is $P\bar{6}m2$, the generators of this space group are threefold rotational symmetry along the z direction, out-of-plane mirror symmetry and in-plane mirror
symmetry.  The in-plane magnetization  preserves the symmetry, and so the in-plane magnetic calculations with SOC show that the energies of -K and K valleys remain almost degenerate\cite{ap}.  The out-of-plane magnetization violates the out-of-plane mirror symmetry,
the SOC effect will remove the degeneracy of -K and K valleys\cite{ap}. To make  $\mathrm{GdI_2}$ monolayer become polarized, an external out-of-plane magnetic field  should be applied. Among $\mathrm{GdX_2}$ (X=F, Cl, Br and I) monolayers,   the
easy axis of  monolayer $\mathrm{GdF_2}$ and $\mathrm{GdCl_2}$ are along the out-of-plane direction\cite{g0,g2}.
For $\mathrm{GdCl_2}$, the K valley is polarized with the valley splitting of 42.3 meV, and the polarized valley is at -K point with valley splitting of 47.6 meV for  $\mathrm{GdF_2}$\cite{g0}.  A compressive strain can decrease
the valley splitting of $\mathrm{GdCl_2}$, and  the polarized valley changes from K point to -K point at about 0.963 strain\cite{g0}.

In light of these  factors about monolayer $\mathrm{GdF_2}$ and $\mathrm{GdCl_2}$  mentioned above,  the Janus monolayer GdClF may be a ferrovalley  material with negligible valley splitting. In consideration of Janus monolayer MoSSe from $\mathrm{MoS_2}$\cite{e1,e2}, the Janus monolayer  GdClF can be synthesized experimentally with similar  synthetic technology based on $\mathrm{GdCl_2}$ monolayer.
The broken  space inversion symmetry
allows GdClF monolayer to become piezoelectric.
In this work,   we  mainly investigate the strain effects on  valley physics and  piezoelectric properties of monolayer GdClF by the first-principles calculations.  The GdClF monolayer is
demonstrated to be a room temperature ferrovalley  material with very little valley splitting.
The compressive strain can enhance splitting of -K valley polarization, while the tensile strain can improve the splitting of  K valley polarization.
From compressive strain to tensile one, the strain can overturn in-plane piezoelectric polarizations.
Our works
provide a potential 2D
materials to achieve polarization of different valleys tuned by strain.

The rest of the paper is organized as follows. In the next
section, we shall give our computational details and methods.
 In  the next few sections,  we shall present structure and stability, electronic structure and valley properties, and piezoelectric properties of Janus monolayer GdClF. Finally, we shall give our conclusions.

\section{Computational detail}
We perform the spin-polarized  first-principles calculations   within density functional theory (DFT)\cite{1},  as implemented in VASP code\cite{pv1,pv2,pv3} within the projector augmented-wave (PAW) method.  The generalized gradient
approximation  of Perdew-Burke-Ernzerhof (PBE-GGA) functional is adopted as the exchange-correlation functional.
The kinetic energy cutoff and total energy  convergence criterion  is set to 500 eV and  $10^{-8}$ eV, respectively.
For optimized  convergence criterion,  less than 0.0001 $\mathrm{eV.{\AA}^{-1}}$  for force on each atom is used for atomic coordinates.
To avoid adjacent interactions, the vacuum
space along the out-of-plane direction  is set to more than 18 $\mathrm{{\AA}}$.
 We use 18$\times$18$\times$1 Monkhorst-Pack k-point mesh to sample the Brillouin zone for calculating electronic structures and elastic properties, and 10$\times$20$\times$1 Monkhorst-Pack k-point mesh for piezoelectric calculations.
To account for the localized nature of 4$f$ orbitals of Gd
atoms, a Hubbard correction $U_{eff}$=4 eV\cite{g2}  is employed within the
rotationally invariant approach proposed by Dudarev et al, and the SOC
is incorporated for investigation of valley properties.

The elastic stiffness tensor  $C_{ij}$  and  piezoelectric stress tensor $e_{ij}$ are calculated by using strain-stress relationship (SSR) and density functional perturbation theory (DFPT) method\cite{pv6}, respectively.
The 2D elastic coefficients $C^{2D}_{ij}$
 and   piezoelectric stress coefficients $e^{2D}_{ij}$
have been renormalized by   $C^{2D}_{ij}$=$L_z$$C^{3D}_{ij}$ and $e^{2D}_{ij}$=$L_z$$e^{3D}_{ij}$, where the $L_z$ is  the length of unit cell along z direction.  The interatomic force constants (IFCs) of monolayer GdClF are calculated  within finite displacement method, and
the 5$\times$5$\times$1 supercell is used with FM ordering. Based on the harmonic IFCs, phonon dispersion spectrum  is attained by the  Phonopy code\cite{pv5}. We use a  40$\times$40 supercell and  $10^7$ loops  to performe the
Monte Carlo (MC) simulations, as implemented in Mcsolver code\cite{mc}.
The ab initio
molecular dynamics (AIMD) simulations  using NVT ensemble are performed with a supercell
of size 4$\times$4$\times$1 for more than
5000 fs with a time step of 1 fs.

\begin{figure}
  \includegraphics[width=8cm]{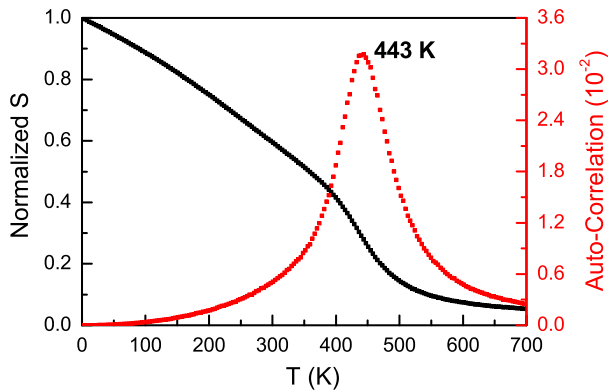}
\caption{(Color online)The normalized magnetic moment (S) and auto-correlation of Janus monolayer  $\mathrm{GdClF}$ as a function of temperature.}\label{tc}
\end{figure}


\begin{figure*}
  \includegraphics[width=12cm]{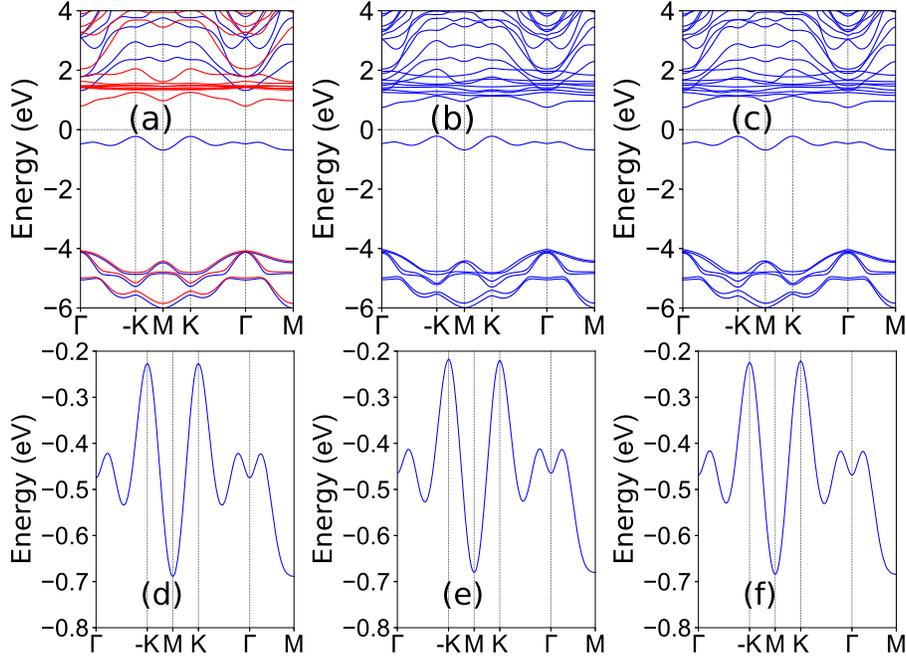}
  \caption{(Color online) The energy  band structures of  Janus  monolayer  $\mathrm{GdClF}$ (a) without SOC; (b) and (c) with SOC for magnetic moment of Gd along the positive and negative z direction, respectively.  The (d), (e) and (f) are  enlarged views of the valence bands near the Fermi level for (a), (b) and (c). The blue (red) lines represent the band structure in the spin-up (spin-down) direction for (a).}\label{band}
\end{figure*}
\begin{figure*}
  \includegraphics[width=15cm]{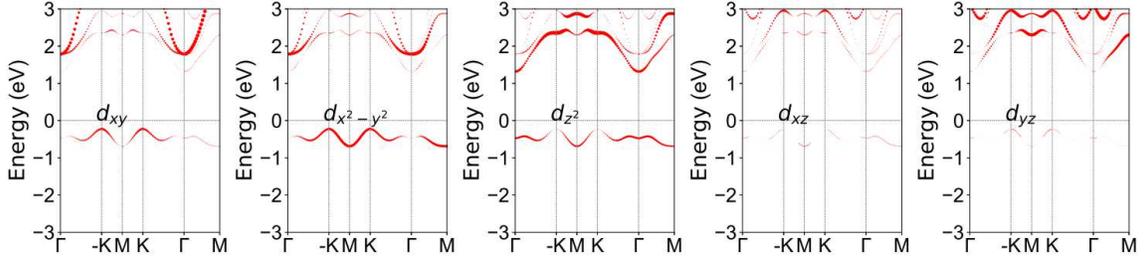}
  \caption{(Color online) For monolayer GdClF, the Gd-$d$-orbital characters of energy bands using GGA for spin-up  direction.  }\label{d-p}
\end{figure*}

\begin{figure*}
  \includegraphics[width=15cm]{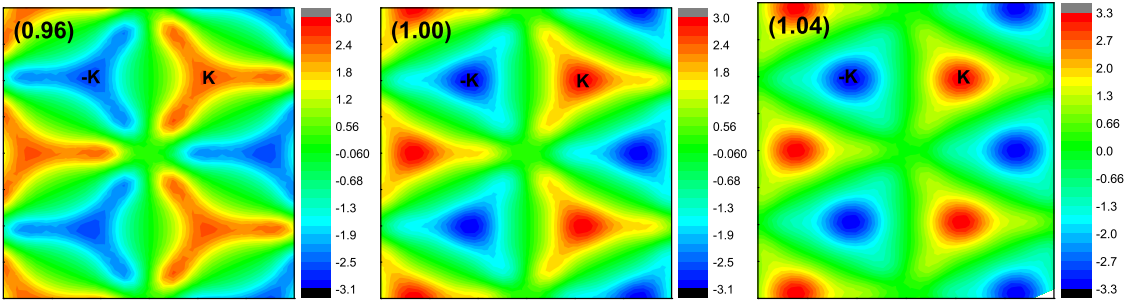}
  \caption{(Color online) Calculated Berry curvature distribution of Janus  monolayer  $\mathrm{GdClF}$  in the 2D Brillouin zone at 0.96, 1.00 and 1.04 strains.  }\label{berry}
\end{figure*}

\begin{figure}
  \includegraphics[width=7cm]{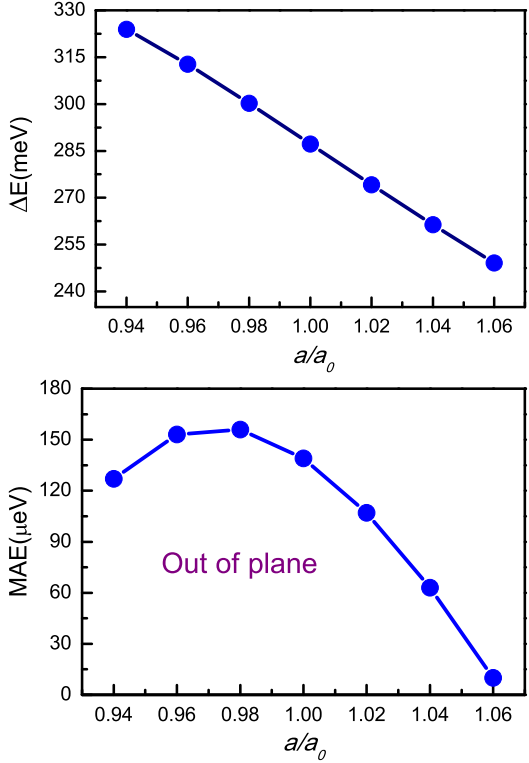}
\caption{(Color online)  (Top) The energy difference between the AFM and FM states ($\Delta E$) and (Bottom)  MAE
as a function of the applied biaxial strain $a/a_0$ for Janus monolayer  $\mathrm{GdClF}$.}\label{s-1}
\end{figure}

\begin{figure*}
  \includegraphics[width=12cm]{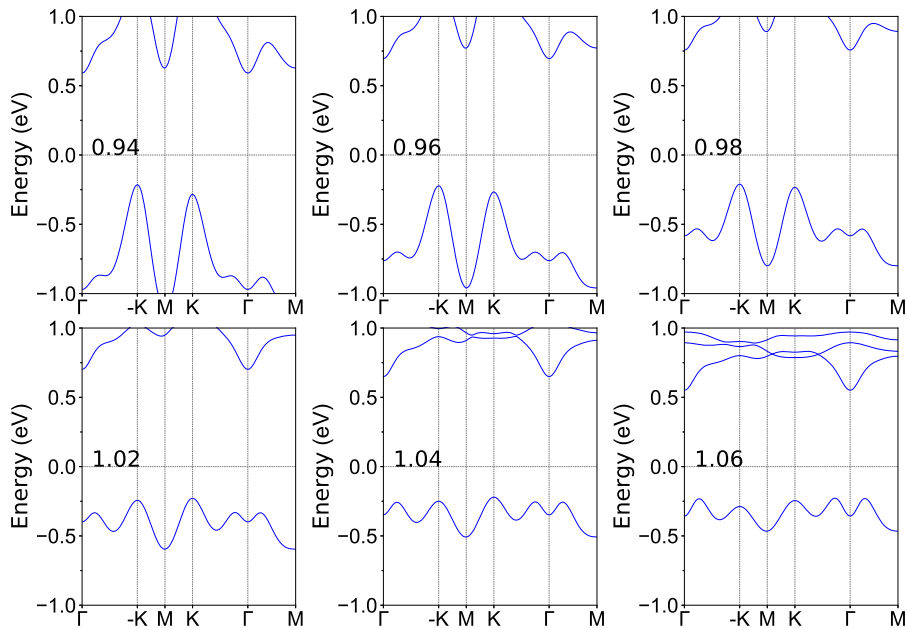}
\caption{(Color online) The strained  energy band structures  of Janus  monolayer  $\mathrm{GdClF}$   with $a/a_0$ from 0.94 to 1.06 by using GGA+SOC.}\label{s-band}
\end{figure*}

\begin{figure}
  \includegraphics[width=7cm]{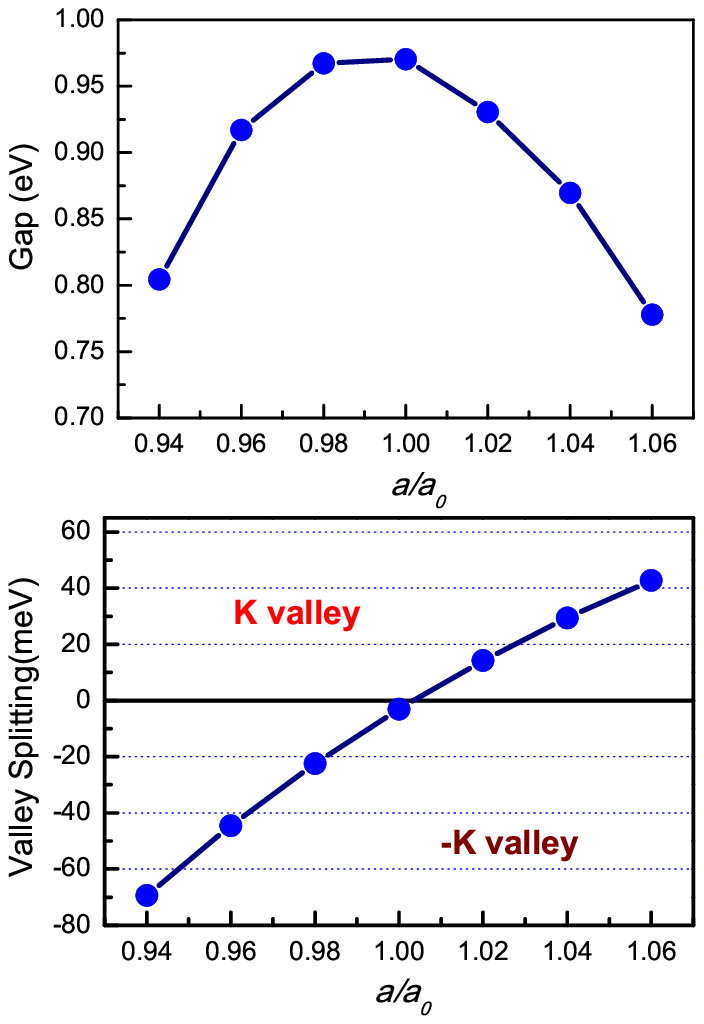}
\caption{(Color online) (Top) The energy band gap (Gap) and  (Bottom) valley splitting
as a function of the applied biaxial strain $a/a_0$ for Janus monolayer  $\mathrm{GdClF}$.}\label{s-2}
\end{figure}

\section{Structure and stability}
 The top and side views of  crystal structures of the Janus GdClF
monolayer are shown in \autoref{st} (a and b). The GdClF monolayer  consists of Cl-Gd-F sandwich layer, and each Gd
atom is surrounded by three Cl and three F atoms forming a triangular prism.
The  Janus monolayer GdClF can be constructed  by  replacing one of two  Cl (F)  layers with F (Cl)  atoms in  $\mathrm{GdCl_2}$ ($\mathrm{GdF_2}$) monolayer. Due to  missing out-of-plane mirror symmetry,  the GdClF monolayer (space group No.156) has lower symmetry compared with $\mathrm{GdCl_2}$/$\mathrm{GdF_2}$ monolayer (space group No.187).
\autoref{st} (d) shows the energies of antiferromagnetic (AFM) and FM states with rectangle supercell (FIG.1 of electronic supplementary information (ESI)) as a function of lattice constants $a$, which implies  that the FM order is the most stable magnetic state. The optimized lattice constants with FM order is 3.61  $\mathrm{{\AA}}$, which is between ones of $\mathrm{GdCl_2}$ and $\mathrm{GdF_2}$\cite{g0,g2}. Due to the different atomic sizes and electronegativities of Cl and F atoms, the GdClF monolayer shows
the  inequivalent Gd-Cl and Gd-F bond lengths (Cl-Gd-Cl and F-Gd-F bond angles), giving rise to a built-in electric field.
The corresponding values are
2.772 $\mathrm{{\AA}}$ and 2.428 $\mathrm{{\AA}}$ (81.27$^{\circ}$ and 96.02$^{\circ}$).

The magnetic anisotropy energy (MAE) can be used to describe magnetic anisotropy of 2D materials, which  plays a important role to realize the long-range magnetic ordering. The  $\mathrm{GdBr_2}$ and $\mathrm{GdI_2}$ monolayers  possess in-plane magnetic anisotropy, but the direction of the easy axis of $\mathrm{GdCl_2}$ and $\mathrm{GdF_2}$ monolayers  are out-of-plane (z-axis)\cite{g0,g1,g2}.
By considering SOC interaction, the MAE is  defined by $E_{MAE}$ = $E_{(100)}$-$E_{(001)}$, and the corresponding value is 139 $\mathrm{\mu eV}$/Gd.
The positive value means that the easy axis of monolayer GdClF is perpendicular to the plane,  possessing PMA, as expected.

The Curie temperature $T_C$ is a key parameter for the practical applications. The $T_C$ of monolayer GdClF is estimated by using the
Wolf  MC algorithm based on the Heisenberg model. The spin Heisenberg Hamiltonian can be expressed  as:
  \begin{equation}\label{pe0-1-1}
H=-J\sum_{i,j}S_i\cdot S_j-A\sum_i(S_i^z)^2
 \end{equation}
in which  $S_i$ and $S_j$ are   the
spin vectors of each atom, and $S_i^z$  is the spin component parallel to the z direction, and $J$ is  the nearest neighbor exchange parameter, and $A$  represents MAE. The energies ($E_{FM}$ and $E_{AFM}$) of FM and AFM
 configurations of monolayer GdClF with rectangle supercell can be written as:
  \begin{equation}\label{pe0-1-2}
E_{FM}=E_0-6J|S|^2-2A|S^2|
 \end{equation}
  \begin{equation}\label{pe0-1-3}
E_{AFM}=E_0+2J|S|^2-2A|S^2|
 \end{equation}
where $E_0$ is the total energy of systems
without magnetic coupling, and  $|S|$ ($|S|$=4
for Gd atoms) is the total spin. The nearest neighbor exchange parameter J is calculated as:
  \begin{equation}\label{pe0-1-3}
J=\frac{E_{AFM}-E_{FM}}{8|S|^2}
 \end{equation}
The calculated $J$ is 2.244 meV based on  $E_{AFM}$-$E_{FM}$=287.236 meV  with rectangle supercell.
 The  normalized magnetic moment and auto-correlation of monolayer GdClF  vs temperature are shown in \autoref{tc}, and the predicted $T_C$ is about 443 K. In ref.\cite{g2}, the calculated $J$ of  $\mathrm{GdCl_2}$ monolayer  is 1.068 meV based on  $E_{AFM}$-$E_{FM}$=136.8 meV  with rhombus primitive cell, which leads to small $T_C$ of 224 K\cite{g2}.

To demonstrate the dynamical stability of GdClF monolayer, the phonon spectra without imaginary frequency modes are shown in \autoref{st} (c).
 The free energy  of Janus  monolayer  $\mathrm{GdClF}$ as a function of simulation time at room
temperature is plotted \autoref{st} (e), along with final structures  after 5 ps. The free energy shows little oscillation as the time increases, and there is no structural destruction or bond broken during the simulation. These mean that monolayer GdClF is dynamically and thermally  stable.
To check
the mechanical stability of monolayer  GdClF,  the elastic
constants is calculated. Using Voigt notation, the independent elastic constants are  $C_{11}$ and $C_{12}$ due to hexagonal symmetry.
The  $C_{11}$ and $C_{12}$ are 62.04 $\mathrm{Nm^{-1}}$ and 17.41 $\mathrm{Nm^{-1}}$,  which satisfy the  Born  criteria of mechanical stability\cite{ela}: $C_{11}$$>$0 and $C_{11}-C_{12}$$>$0,   confirming the mechanical stability of monolayer  GdClF. The  monolayer  GdClF is  mechanically isotropic, so its  Young¡¯s moduli $C^{2D}$, shear modulus $G^{2D}$ and Poisson's ratios $\nu^{2D}$ can simply be written as \cite{ela}:
\begin{equation}\label{e1}
C^{2D}=\frac{C_{11}^2-C_{12}^2}{C_{11}}
\end{equation}
\begin{equation}\label{e1}
G^{2D}=C_{66}=\frac{C_{11}-C_{12}}{2}
\end{equation}
\begin{equation}\label{e1}
\nu^{2D}=\frac{C_{12}}{C_{11}}
\end{equation}
The calculated results show  that Young's moduli $C_{2D}$,  shear modulus $G_{2D}$ and Poisson's ratio $\nu$  of GdClF monolayer are 57.15 $\mathrm{Nm^{-1}}$, 22.32  $\mathrm{Nm^{-1}}$ and 0.281, respectively.  The  $C_{2D}$ is
less than those  of many 2D materials\cite{gra,q7-0,q7-1,q7-2,q7-3,q7-4}, which means  that GdClF monolayer can be easily tuned by strain.

\section{Electronic structure and valley properties}
At absence of  SOC, the spin-polarized band structure of monolayer  GdClF is shown in \autoref{band} (a,d).
It is found that GdClF monolayer is an indirect band gap semiconductor with VBM and conduction band minimum (CBM)  provided by the spin-up and spin-dn, and they locate at the K and $\Gamma$  symmetry points  with the gap value of 1.025 eV. This means that GdClF monolayer is a  bipolar magnetic semiconductor with 100 \% spin-polarized currents with inverse spin-polarization direction for electron or hole doping. The monolayer GdClF may be  a potential
ferrovalley material due to degenerate energy
extremes of  K and -K high-symmetry points in the valence band (\autoref{band} (d)). The Gd-$d$-orbital characters of energy bands for spin-up  direction are plotted in \autoref{d-p}.  It was clearly seen  that
 the VBM is predominantly
contributed by $d_{x^2-y^2}$ and $d_{xy}$ orbitals, which is very important to produce valley polarization\cite{q10}, when the SOC is included.

With SOC, the  energy band structures of  monolayer  GdClF  with  magnetic moment of Gd along the positive and negative out-of-plane direction  are also plotted in \autoref{band} (b,e and c,f). The SOC effect can induce valley polarization in uppermost valence band (UVB), which is due to  the intra-atomic interaction $\hat{L}\cdot\hat{S}$ ($\hat{H}^0_{SOC}+\hat{H}^1_{SOC}=\lambda\hat{L}\cdot\hat{S}$), where  $\hat{L}$ and $\hat{S}$ represent the orbital angular moment and spin angular moment, respectively.  The $\hat{H}^0_{SOC}$
represents the interaction
between the same spin states, which dominates the  $\hat{L}\cdot\hat{S}$.
 Due to the magnetic exchange interaction, the
opposite spin states interaction ($\hat{H}^1_{SOC}$) can be  ignored. The  $\hat{H}^0_{SOC}$ can be expressed as\cite{q10,q18}:
\begin{equation}\label{e1}
\hat{H}^0_{SOC}=\lambda\hat{S}_{z^`}(\hat{L}_z cos\theta+\frac{1}{2}\hat{L}_{+}e^{-i\phi}sin\theta+\frac{1}{2}\hat{L}_{-}e^{+i\phi}sin\theta)
\end{equation}
 The $\theta$ and $\phi$ are the polar angles of spin orientation, which are defined in FIG.1 of ESI. Because the MAE is out-of-plane ($\theta=0$$^{\circ}$),  $\hat{H}^0_{SOC}$ can be simplified as:
\begin{equation}\label{m1}
\hat{H}^0_{SOC}=\alpha \hat{L}_z
\end{equation}
 The \autoref{m1} suggests  that the
Hamiltonian $\hat{H}^0_{SOC}$ relies on the orbital angular momentum $\hat{L}_z$. The group symmetry of the wave vector at K and -K points is $C_{3h}$, and  the basis functions can be expressed as\cite{q10,q18}:
\begin{equation}\label{m2}
|\phi^\tau_\nu>=\sqrt{\frac{1}{2}}(|d_{x^2-y^2}>+i\tau|d_{xy}>)
\end{equation}
where the subscript $\nu$ and $\tau$ represent valence bands and the valley index ($\tau=\pm1$).  Based on the
basis functions, the energy levels of the VBM  at K
and -K valleys can be expressed as:
\begin{equation}\label{m3}
E^\tau_\nu=<\phi^\tau_\nu|\hat{H}^0_{SOC}|\phi^\tau_\nu>
\end{equation}
Then, the valence  energy difference $\Delta E_\nu$
 of valley at -K and K points (valley splitting) can be expressed as:
\begin{equation}\label{m4}
\Delta E_\nu=E^{K}_\nu-E^{-K}_\nu=4\alpha
\end{equation}

The DFT calculated results show that the $\Delta E_\nu$ is only -3.1 meV, which means that  the energy of K valley state
is lower than one of -K valley (\autoref{band} (e)). By applying an external magnetic
field, the  valley polarization of monolayer  GdClF can change from -K point to K point.
The reversing  magnetization
of Gd atoms can simultaneously flip  the spin and valley polarization, and the energy of K valley will become higher than one  of -K valley (\autoref{band} (f)),  which suggests
that the valley polarization can be tuned by the magnetization orientation. This can be explained by $\Delta E_\nu=4\alpha cos\theta$\cite{q18}, when considering the magnetization orientation ($\theta$ changes from 0$^{\circ}$ to 180$^{\circ}$). When the magnetization orientation is along in-plane direction ($\theta$=90$^{\circ}$), the valley splitting will disappear ($\Delta E_\nu$=0). So, it is a key factor for valley application to possess PMA.
 The UVB related with valley polarization  is separated well  from other energy bands, which is very useful to  manipulate valley states.

 The electronic transport properties and various Hall effects are related with  Berry curvature. Due to broken space inversion symmetry and exchange
interaction for hexagonal systems, the carriers will attain  a nonzero un-equivalent
Berry curvature at K and -K valleys, which can induce AVHE\cite{q10}.
 The Berry curvature of GdClF monolayer
is calculated directly from the calculated
wave functions  based on Fukui's
method\cite{bm},  as implemented in the VASPBERRY code.
The calculated Berry curvature distribution in the 2D Brillouin zone  with SOC for magnetic moment of Gd along the positive  z direction  is plotted in \autoref{berry}. It is found that   the  Berry curvatures of K and -K
valleys have opposite signs, and the absolute values  are different,  giving rise to the typical valley contrasting properties.

For  $\mathrm{GdCl_2}$ monolayer, a compressive strain can reduce
the valley splitting, and  the valley splitting will change sign at about 0.963 strain\cite{g0}. The valley splitting $\Delta E_\nu$ of unstrained GdClF is very small, which provides a possibility to achieve valley polarization at different valley points by tensile and compressive strains.
We use $a/a_0$ to simulate the biaxial strain, where   $a$ and $a_0$ are the strained and  unstrained lattice constants. The $a/a_0$$<$1 means the compressive strain, and the $a/a_0$$>$1 represents the tensile strain. Firstly, we determine the magnetic ground state of strained GdClF monolayer by
the energy differences of   AFM  with respect to  FM state  vs $a/a_0$, which is shown in \autoref{s-1}.
In considered strain range, the DFT calculated results show that the energy difference  is always positive, and  monotonically decreases with $a/a_0$ from 0.94 to 1.06. This confirms the FM ground state of strained monolayer GdClF  in considered strain range. It is found that  the
compressive strain  can strengthen  the FM coupling between Gd atoms, while the tensile strain can reduce the FM coupling.
Another factor is to confirm PMA of strained GdClF monolayer to possess stable  long-range
magnetic ordering without external field.  The MAE as a function of $a/a_0$ is also plotted in \autoref{s-1}, which  shows a decrease with increasing $a/a_0$. However, the MAE is always  positive, which means that the easy axis of strained monolayer GdClF is always along z direction.

The energy band structures of strained monolayer GdClF by using GGA+SOC are plotted in \autoref{s-band},
and the corresponding energy band gaps  are shown in \autoref{s-2}. The strained  monolayer GdClF is always an indirect gap semiconductor, and the CBM  maintains at $\Gamma$ point. For VBM, the tensile strain (about 1.06) can change it from K/-K point to one point along (K/-K)-$\Gamma$ path. The energy band gap  firstly increases, and then decreases, when the strain changes from compressive one to tensile one. Similar phenomenon can be found in many 2D materials\cite{gsd1,gsd2}. The valley splitting vs $a/a_0$  is also plotted in \autoref{s-2}. The compressive strain can make the absolute value of  valley splitting become large, inducing -K valley polarization.  However, the increasing  tensile strain enhances the valley splitting, giving rise to K valley polarization. At moderate 0.96 and 1.04 strain, the corresponding valley splitting are -44.5 meV and 29.4 meV, which are larger than
the thermal energy of  room temperature (25 meV). This is very important for room-temperature electronic device.  The Berry curvature distributions of monolayer  GdClF  at 0.96 and 1.04 strains by using GGA+SOC are plotted in \autoref{berry}. With increasing $a/a_0$,  the Berry curvatures  (absolute value) of two
valleys increase slightly. Due to small Young's moduli $C_{2D}$ of GdClF monolayer, the  0.96 and 1.04 strains can be easily achieved  experimentally.
These confirm the achievement of different valley polarization by strain.
\begin{table}
\centering \caption{ For  monolayer  $\mathrm{GdF_2}$\cite{g0}, GdClF and   $\mathrm{GdCl_2}$\cite{g0},   the elastic constants $C_{ij}$ in $\mathrm{Nm^{-1}}$, the piezoelectric stress coefficients $e_{11}$/$e_{31}$ in $10^{-10}$ C/m, and the piezoelectric strain coefficients $d_{11}$/$d_{31}$ in pm/V. }\label{tab}
  \begin{tabular*}{0.48\textwidth}{@{\extracolsep{\fill}}ccccccc}
  \hline\hline
Name& $C_{11}$ &  $C_{12}$& $e_{11}$& $e_{31}$& $d_{11}$&$d_{31}$\\\hline\hline
 $\mathrm{GdF_2}$&73.87&19.61&0.317&-&0.584&-\\\hline
 GdClF& 62.04&17.41&-0.260&-0.181&-0.583&-0.228 \\\hline
 $\mathrm{GdCl_2}$&45.95&13.53&-0.878&-&-2.708&-\\\hline\hline
\end{tabular*}
\end{table}

\section{Piezoelectric properties}
 The  ferrovalley materials  should break  inversion symmetry, which means that they are  piezoelectric.
The piezoelectric effects of a material can be described by  third-rank piezoelectric stress tensor  $e_{ijk}$ and strain tensor $d_{ijk}$, which can be attained from the sum of ionic
and electronic parts.  They  are defined as:
 \begin{equation}\label{pe0}
      e_{ijk}=\frac{\partial P_i}{\partial \varepsilon_{jk}}=e_{ijk}^{elc}+e_{ijk}^{ion}
 \end{equation}
and
 \begin{equation}\label{pe0-1}
   d_{ijk}=\frac{\partial P_i}{\partial \sigma_{jk}}=d_{ijk}^{elc}+d_{ijk}^{ion}
 \end{equation}
Where $P_i$, $\varepsilon_{jk}$ and $\sigma_{jk}$ are polarization vector, strain and stress, respectively.
The $e_{ijk}^{elc}$/$d_{ijk}^{elc}$ is clamped-ion piezoelectric coefficients, not including ionic contribution. The  $e_{ijk}$/$d_{ijk}$ is relax-ion piezoelectric coefficients as a realistic result.

For GdClF monolayer, it not only lacks  inversion symmetry, but also breaks vertical mirror symmetry. This is different from
monolayer $\mathrm{GdCl_2}$, whose vertical mirror symmetry still holds. These mean that  the piezoelectric coefficients  $e_{11}$/$d_{11}$  and $e_{31}$/$d_{31}$ of GdClF monolayer are  nonzero. This is the same with ones of Janus monolayer MoSSe\cite{q7-0}. For 2D materials,  by using  Voigt notation with only considering  the in-plane strain and stress\cite{q7-0,q7-1,q7-2,q7-3,q7-4}, the  piezoelectric stress, strain and elastic tensors  can be reduced into:
 \begin{equation}\label{pe1-1}
 e=\left(
    \begin{array}{ccc}
      e_{11} & -e_{11} & 0 \\
     0 & 0 & -e_{11} \\
      e_{31} & e_{31} & 0 \\
    \end{array}
  \right)
    \end{equation}

  \begin{equation}\label{pe1-2}
  d= \left(
    \begin{array}{ccc}
      d_{11} & -d_{11} & 0 \\
      0 & 0 & -2d_{11} \\
      d_{31} & d_{31} &0 \\
    \end{array}
  \right)
\end{equation}

 \begin{equation}\label{pe1-3}
   C=\left(
    \begin{array}{ccc}
      C_{11} & C_{12} & 0 \\
     C_{12} & C_{11} &0 \\
      0 & 0 & (C_{11}-C_{12})/2 \\
    \end{array}
  \right)
\end{equation}
  With an imposed uniaxial in-plane strain,   the in-plane and out of plane piezoelectric polarization ($e_{11}$/$d_{11}$ and $e_{31}$/$d_{31}$$\neq$0) can be induced. However, when  a  biaxial in-plane strain is apllied,  the
in-plane piezoelectric response will be suppressed($e_{11}$/$d_{11}$=0), but the out of plane piezoelectric polarization still can be induced ($e_{31}$/$d_{31}$$\neq$0).
The  $d_{11}$ and $d_{31}$ can be calculated  by $e_{ik}=d_{ij}C_{jk}$:
\begin{equation}\label{pe2}
    d_{11}=\frac{e_{11}}{C_{11}-C_{12}}~~~and~~~d_{31}=\frac{e_{31}}{C_{11}+C_{12}}
\end{equation}

\begin{figure}
  \includegraphics[width=8cm]{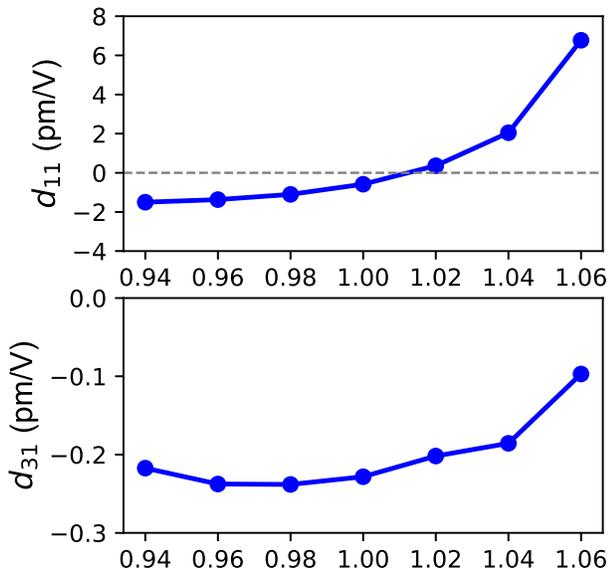}
\caption{(Color online) For monolayer  GdClF, the piezoelectric strain coefficient ($d_{11}$ and $d_{31}$) as a function of $a/a_0$.}\label{d-s}
\end{figure}

The orthorhombic supercell is used  to calculate the  $e_{11}$/$e_{31}$  of monolayer GdClF by using DFPT method.
The calculated $e_{11}$/$e_{31}$ is -0.260$\times$$10^{-10}$/ -0.181$\times$$10^{-10}$ C/m  with ionic part -1.558$\times$$10^{-10}$/-0.276$\times$$10^{-10}$ C/m  and electronic part 1.298$\times$$10^{-10}$/0.095$\times$$10^{-10}$ C/m.
The electronic and ionic contributions of both $e_{11}$ and $e_{31}$ have  opposite signs, and   the ionic part
 dominates the  piezoelectricity.
Based on \autoref{pe2}, the calculated  $d_{11}$ and $d_{31}$  are  -0.583 and -0.228 pm/V.
The  $\mathrm{GdX_2}$ (X= Cl and F) monolayers  have been predicted, and they possess only in-plane piezoelectricity\cite{g0}.
The data related with elastic and piezoelectric properties of monolayer  $\mathrm{GdF_2}$, GdClF and   $\mathrm{GdCl_2}$  are summarized in
\autoref{tab}. It is found that the $d_{11}$ (absolute value) of  monolayer GdClF is smaller than one of  $\mathrm{GdCl_2}$, and is almost the same with one of $\mathrm{GdF_2}$.

The strain effects on piezoelectric properties of monolayer  GdClF are investigated.  The elastic constants  ($C_{11}$, $C_{12}$,  $C_{11}$-$C_{12}$ and $C_{11}$+$C_{12}$) and  piezoelectric  stress  coefficients  ($e_{11}$ and $e_{31}$) along  the ionic  and electronic contributions of monolayer  GdClF as a function of $a/a_0$ are plotted in FIG.2 and FIG.3 of ESI. The $d_{11}$ and $d_{31}$ vs $a/a_0$ are plotted in \autoref{d-s}.
In the considered strain range, all strained  monolayer GdClF are  mechanically stable, because the calculated elastic constants all satisfy  the mechanical stability criteria\cite{ela}. With $a/a_0$ from 0.94 to 1.06, the $d_{11}$ changes from negative value to positive one at about 1.015 strain, which mainly depends on $e_{11}$. When $a/a_0$ is less than about 1.05, the ionic  and electronic parts have opposite contributions to $e_{11}$. At 1.06, a very large $d_{11}$ (6.77 pm/V) is observed, which is mainly  due to superimposed contribution from  the ionic  and electronic parts.
The  $d_{31}$ (absolute value) shows a decreasing trend with strain from compressive one to tensile one.
At 0.96 and 1.04 strains,  the $d_{11}$ of  monolayer GdClF are -1.37 pm/V  and 2.05 pm/V, which provides the possibility for the combination of valley properties and piezoelectricity.

\section{Conclusion}
In summary, we present a model for
valleytronics to induce   polarization of different valley points. We demonstrate our ideas by a concrete 2D material GdClF by  first-principles calculations.   Monolayer  GdClF  is predicted to be a FM
semiconductor with a pair of valleys locating at the K and -K
points, and possess PMA and high $T_C$. Due to the intrinsic   magnetic
interaction and SOC, the valley splitting  between K and -K valleys can be induced, but the value is very little.
By imposed increasing  compressive strain,  the -K valley polarization is produced with enhanced valley splitting.
On the contrary, the valley splitting is improved for K valley polarization, when the tensile strain increases.
Additionally, the piezoelectric properties of monolayer  GdClF are also studied. It is found that the strain can reverse  the direction of in-plane piezoelectric polarizations, when the strain changes from compressive one to tensile one.
Our works  predict  Janus monolayer GdClF with
the combination of valley properties and piezoelectricity. Moreover, the valley polarization and in-plane piezoelectric polarizations can be effectively tuned by strain. These offer  a great potential application in next-generation   multifunctional devices.

\begin{acknowledgments}
This work is supported by Natural Science Basis Research Plan in Shaanxi Province of China  (2021JM-456). We are grateful to the Advanced Analysis and Computation Center of China University of Mining and Technology (CUMT) for the award of CPU hours and WIEN2k/VASP software to accomplish this work.
\end{acknowledgments}

\end{document}